\documentclass[11pt,a4paper,english,preprintnumbers,amsmath,amssymb,nofootinbib]{revtex4}
\usepackage{lmodern}
\usepackage{lmodern}
\usepackage[T1]{fontenc}
\usepackage[latin9]{inputenc}
\usepackage{color}
\usepackage{babel}
\usepackage{amsmath}
\usepackage{amssymb}
\usepackage{esint}
\usepackage[unicode=true,pdfusetitle,
 bookmarks=true,bookmarksnumbered=false,bookmarksopen=false,
 breaklinks=false,pdfborder={0 0 1},backref=false,colorlinks=true]
 {hyperref}
\hypersetup{
 linkcolor=blue,citecolor=blue}

\makeatletter

\pdfpageheight\paperheight
\pdfpagewidth\paperwidth

\@ifundefined{textcolor}{}
{%
 \definecolor{BLACK}{gray}{0}
 \definecolor{WHITE}{gray}{1}
 \definecolor{RED}{rgb}{1,0,0}
 \definecolor{GREEN}{rgb}{0,1,0}
 \definecolor{BLUE}{rgb}{0,0,1}
 \definecolor{CYAN}{cmyk}{1,0,0,0}
 \definecolor{MAGENTA}{cmyk}{0,1,0,0}
 \definecolor{YELLOW}{cmyk}{0,0,1,0}
}

\usepackage{babel}

\@ifundefined{definecolor}{color}{}

\usepackage{latexsym}\usepackage{bm}

\makeatother

\begin{document}
\title{Is there a novel Einstein-Gauss-Bonnet theory in four dimensions?}
\author{Metin Gürses}
\email{gurses@fen.bilkent.edu.tr}

\affiliation{{\small{}{}Department of Mathematics, Faculty of Sciences}\\
 {\small{}{}Bilkent University, 06800 Ankara, Turkey}}
\author{Tahsin Ça\u{g}r\i{} \c{S}i\c{s}man}
\email{tahsin.c.sisman@gmail.com}

\affiliation{Department of Astronautical Engineering,\\
 University of Turkish Aeronautical Association, 06790 Ankara, Turkey}
\author{Bayram Tekin}
\email{btekin@metu.edu.tr}

\affiliation{Department of Physics, Middle East Technical University, 06800 Ankara,
Turkey}
\begin{abstract}
No! We show that the field equations of Einstein-Gauss-Bonnet theory
defined in generic $D>4$ dimensions split into two parts one of which
always remains higher dimensional, and hence the theory does not have
a non-trivial limit to $D=4$. Therefore, the recently introduced
four-dimensional, novel, Einstein-Gauss-Bonnet theory does not admit
an \emph{intrinsically} four-dimensional definition as such it does
not exist in four dimensions. The solutions (the spacetime, the metric)
always remain $D>4$ dimensional. As there is no canonical choice
of 4 spacetime dimensions out of $D$ dimensions for generic metrics,
the theory is not well defined in four dimensions. 
\end{abstract}
\maketitle
\vspace{0.1cm}

\section{Introduction}

Recently a \textit{four}\emph{-dimensional} Einstein-Gauss-Bonnet
theory that is claimed to propagate only a massless spin-2 graviton
was introduced as a limit in \cite{GlavanLin} with the action

\begin{align}
I & =\int d^{D}x\sqrt{-g}\left[\frac{1}{\kappa}\left(R-2\Lambda_{0}\right)+\frac{\alpha}{D-4}\left(R_{\alpha\beta\rho\sigma}R^{\alpha\beta\rho\sigma}-4R_{\alpha\beta}R^{\alpha\beta}+R^{2}\right)\right],\label{eq:Action}
\end{align}
of which the field equations are \cite{DeserTekin,DeserTekinPRD}
\begin{equation}
\frac{1}{\kappa}\left(R_{\mu\nu}-\frac{1}{2}g_{\mu\nu}R+\Lambda_{0}g_{\mu\nu}\right)+\frac{\alpha}{D-4}{\cal H}_{\mu\nu}=0\label{eq:EGB}
\end{equation}
where the ``Gauss-Bonnet tensor'' reads 
\begin{equation}
{\cal H}_{\mu\nu}=2\left[RR_{\mu\nu}-2R_{\mu\alpha\nu\beta}R^{\alpha\beta}+R_{\mu\alpha\beta\sigma}R_{\nu}^{\phantom{\nu}\alpha\beta\sigma}-2R_{\mu\alpha}R_{\nu}^{\alpha}-\frac{1}{4}g_{\mu\nu}\left(R_{\alpha\beta\rho\sigma}R^{\alpha\beta\rho\sigma}-4R_{\alpha\beta}R^{\alpha\beta}+R^{2}\right)\right].\label{eq:EoM}
\end{equation}
For $D>4$, (\ref{eq:EGB}) is the well-known Einstein-Gauss-Bonnet
theory which has been studied in the literature in great detail. On
the other hand, for $D=4$, the ${\cal H}_{\mu\nu}$ tensor vanishes
\textit{identically} and hence, as per common knowledge, the field
equations (\ref{eq:EGB}) reduces to the Einstein's theory. This is
because in four dimensions, the Gauss-Bonnet combination $\mathcal{G}\,:=R_{\alpha\beta\rho\sigma}R^{\alpha\beta\rho\sigma}-4R_{\alpha\beta}R^{\alpha\beta}+R^{2}$
can be written as $\mathcal{G}=\epsilon_{\mu\nu\alpha\beta}\epsilon^{\mu\nu\sigma\rho}R^{\alpha\beta\gamma\lambda}R_{\gamma\lambda\sigma\rho}$
and yields a topological action, i.e. the Euler number which is independent
of the metric $g_{\mu\nu}$. This was the state of affairs until the
paper \cite{GlavanLin} implicitly asked the question ``how does
the ${\cal H}_{\mu\nu}$ tensor go to zero as $D\rightarrow4$?''.
The answer is very interesting: because if it goes to zero in the
following way 
\begin{equation}
{\cal H}_{\mu\nu}=\left(D-4\right)\mathcal{Y}_{\mu\nu},\label{eq:Y}
\end{equation}
where $\mathcal{Y}_{\mu\nu}$ is a new tensor to be found, then the
authors of \cite{GlavanLin} argue that in the $D\rightarrow4$ limit,
the field equations (\ref{eq:EGB}) define a four-dimensional theory
in the limit. So namely, the suggested four-dimensional theory would
be the following theory in source-free case: 
\begin{equation}
\lim_{D\rightarrow4}\left[\frac{1}{\kappa}\left(R_{\mu\nu}-\frac{1}{2}g_{\mu\nu}R+\Lambda_{0}g_{\mu\nu}\right)+\frac{\alpha}{D-4}{\cal H}_{\mu\nu}\right]=0.\label{eq:EGBlimit}
\end{equation}

Let us try to understand what the suggested theory is. As there is
no intrinsically defined four-dimensional covariant tensor that the
Gauss-Bonnet tensor reduces to; namely, $\mathcal{Y}_{\mu\nu}$ in
(\ref{eq:Y}) does not exist as guaranteed by the Lovelock theorem
\cite{Lovelock1,Lovelock2,Lanczos} and as will be shown below, the
theory must be defined as a limit. Thus, to compute anything in this
theory, say the perturbative particle content, the maximally symmetric
vacua, the black hole solutions, or any solution with or without a
symmetry, one must do the computation in $D$ dimensions and than
take the $D\rightarrow4$ limit. Surely, for some components of the
the metric such as the spherically symmetric metric, due to the nature
of the the Gauss-Bonnet tensor, this limit might make sense. But,
at the level of the solutions, namely at the level of the full metric,
this limit makes no sense at all. For example, assume that there is
a solution to the theory given locally with the $D$ dimensional metric
$g_{\mu\nu}$ say which has no isometries. Then, as we need to take
the $D\rightarrow4$ limit, which dimensions or coordinates do we
dispose of, is there a canonical prescription? The answer is no! Even
for spherically symmetric solutions of Boulware and Deser, \cite{BD}
studied so far, we do not have the right to dispose any dimension
we choose.

What we have just stated is actually at the foundations of defining
a gravity theory in the Riemannian geometry context. The Riemannian
geometry depends on the number of dimensions, in defining a classical
gravity theory based on geometry one first fixes the number of dimensions
to be some $D$; and as this number changes, the theory changes. There
is no sensible limiting procedure as defined by (\ref{eq:EGBlimit});
there is of course compactification, dimensional reduction \emph{etc}
where all the spacetime dimensions still survive albeit not in equal
sizes generically.

The layout of the paper is as follows: In Section II, we recast the
$D$-dimensional Gauss-Bonnet tensor using the Weyl tensor in such
a way that it naturally splits into two parts. One part has a formal
$D\rightarrow4$ limit, while the other part is always higher dimensional.
In Section III, we give another proof that the theory is non-trivial
only for $D>4$ using the first-order formalism with the vielbein
and the spin-connection. In Section IV, we give a an explicit example
in the form of a direct-product metric where the role of the higher
dimensional part is apparent.

\section{$D\rightarrow4$ Limit Of The Field Equations}

To further lay out our ideas, and to show that there is no four-dimensional
definition of the theory, let us recast the Gauss-Bonnet tensor, in
such a way that we can see the limiting behaviors. For this purpose,
the Weyl tensor, 
\begin{equation}
C_{\mu\alpha\nu\beta}=R_{\mu\alpha\nu\beta}-\frac{2}{\left(D-2\right)}\left(g_{\mu[\nu}R_{\beta]\alpha}-g_{\alpha[\nu}R_{\beta]\mu}\right)+\frac{2}{\left(D-1\right)\left(D-2\right)}Rg_{\mu[\nu}g_{\beta]\alpha},
\end{equation}
becomes rather useful. Using Appendix A of \cite{UnitaryBIinD}, the
Gauss-Bonnet tensor in $D$ dimensions can be split as 
\begin{equation}
{\cal H}_{\mu\nu}=2\left(\mathcal{L}_{\mu\nu}+\mathcal{Z}_{\mu\nu}\right),\label{eq:H_split}
\end{equation}
where the first term does not have an explicit coefficient related
to the number of dimensions and is given as 
\begin{equation}
\mathcal{L}_{\mu\nu}:=C_{\mu\alpha\beta\sigma}C_{\nu}^{\phantom{\nu}\alpha\beta\sigma}-\frac{1}{4}g_{\mu\nu}C_{\alpha\beta\rho\sigma}C^{\alpha\beta\rho\sigma},\label{eq:L}
\end{equation}
which we shall name as the Lanczos-Bach tensor, and the other part
carries explicit coefficients regarding the number of dimensions:
\begin{align}
\mathcal{Z_{\mu\nu}}:= & \frac{\left(D-4\right)\left(D-3\right)}{\left(D-1\right)\left(D-2\right)}\Biggl[-\frac{2\left(D-1\right)}{\left(D-3\right)}C_{\mu\rho\nu\sigma}R^{\rho\sigma}-\frac{2\left(D-1\right)}{\left(D-2\right)}R_{\mu\rho}R_{\nu}^{\rho}+\frac{D}{\left(D-2\right)}R_{\mu\nu}R\nonumber \\
 & \phantom{\frac{\left(D-4\right)\left(D-3\right)}{\left(D-1\right)\left(D-2\right)}\Biggl[}+\frac{1}{\left(D-2\right)}g_{\mu\nu}\left(\left(D-1\right)R_{\rho\sigma}R^{\rho\sigma}-\frac{D+2}{4}R^{2}\right)\Biggr],\label{eq:Z}
\end{align}
where we kept all the factors to see how the limiting procedure might
work. With the $2/\left(D-4\right)$ factor, the second part nicely
reduces to a tensor $\mathcal{S}_{\mu\nu}$ which is finite in the
$D\rightarrow4$ limit; 
\begin{equation}
\mathcal{S}_{\mu\nu}:=\frac{2}{D-4}\mathcal{Z}_{\mu\nu}.\label{eq:S}
\end{equation}
But, the first part is rather non-trivial. In $D=4$ dimensions we
have the Lanczos-Bach identity \cite{Bach,Lanczos} for any smooth
metric; 
\begin{equation}
C_{\mu\alpha\beta\sigma}C_{\nu}^{\phantom{\nu}\alpha\beta\sigma}=\frac{1}{4}g_{\mu\nu}C_{\alpha\beta\rho\sigma}C^{\alpha\beta\rho\sigma}\qquad\text{for all metrics in }D=4.
\end{equation}
Thus, a cursory look might suggest that one might naively assume the
Lanczos-Bach identity in four dimensions and set $\mathcal{L}_{\mu\nu}=0$
in the $D\rightarrow4$ limit yielding a finite intrinsically four
dimensional description of the Gauss-Bonnet tensor as 
\begin{equation}
\lim_{D\rightarrow4}\left(\frac{1}{D-4}{\cal H}_{\mu\nu}\right)=\frac{1}{3}\left[-6C_{\mu\rho\nu\sigma}R^{\rho\sigma}-3R_{\mu\rho}R_{\nu}^{\rho}+2R_{\mu\nu}R+\frac{3}{2}g_{\mu\nu}\left(R_{\rho\sigma}R^{\rho\sigma}-\frac{1}{2}R^{2}\right)\right].
\end{equation}
where the right-hand side is $\mathcal{S}_{\mu\nu}$, given in (\ref{eq:S}),
calculated at $D=4$. But this is a red-herring! The $\mathcal{H}$
tensor or the $\mathcal{S}$ tensor does not obey the Bianchi identity
\begin{equation}
\nabla^{\mu}\mathcal{S}_{\mu\nu}\ne0.
\end{equation}
Therefore, without further assumptions, it cannot be used in the description
of a four dimensional theory. Then, this begs the question: How does
the $\mathcal{L}_{\mu\nu}$ tensor go to zero in the $D\rightarrow4$
limit, that is 
\begin{equation}
\lim_{D\rightarrow4}\left[\frac{1}{D-4}\left(C_{\mu\alpha\beta\sigma}C_{\nu}^{\phantom{\nu}\alpha\beta\sigma}-\frac{1}{4}g_{\mu\nu}C_{\alpha\beta\rho\sigma}C^{\alpha\beta\rho\sigma}\right)\right]=?
\end{equation}
To save the Bianchi identity, $\mathcal{L}_{\mu\nu}$ should have
the form 
\begin{equation}
\frac{2}{D-4}\mathcal{L}_{\mu\nu}=\mathcal{T}_{\mu\nu}\qquad\mbox{for }D\ne4.
\end{equation}
If this is the case, then there is a discontinuity for the Gauss-Bonnet
tensor as 
\begin{equation}
\frac{1}{D-4}{\cal H}_{\mu\nu}=\begin{cases}
\mathcal{T}_{\mu\nu}+\mathcal{S}_{\mu\nu}, & \mbox{for }D\ne4,\\
\frac{0}{0}, & \mbox{for }D=4.
\end{cases}
\end{equation}
Then, in the $D\rightarrow4$ limit, the Gauss-Bonnet tensor with
an $\alpha/D-4$ factor becomes 
\begin{equation}
\lim_{D\rightarrow4}\left(\frac{1}{D-4}{\cal H}_{\mu\nu}\right)=\mathcal{T}_{\mu\nu}+\mathcal{S}_{\mu\nu},\label{eq:Limit_EoM}
\end{equation}
that is the Gauss-Bonnet tensor is not continuous in $D$ at $D=4$.
This discontinuity in the Gauss-Bonnet tensor introduces a problem:
Let $g_{\mu\nu}^{D}$ be the solution of the field equations for $D>4$,
and $g_{\mu\nu}^{\lim}$ is the solution of the limiting field equations
(\ref{eq:Limit_EoM}); then\textit{ 
\begin{equation}
{\displaystyle \lim_{D\to4}g_{\mu\nu}^{D}\ne g_{\mu\nu}^{\lim}},
\end{equation}
}\textit{\emph{in general.}}

Incidentally, the $\mathcal{L}_{\mu\nu}$ tensor is related to the
trace of the $D$ dimensional extension of the Bel-Robinson tensor
given in \cite{Senovilla} which reads 
\begin{eqnarray}
\mathcal{B}_{\alpha\beta\lambda\mu} & = & C_{\alpha\rho\lambda\sigma}C_{\beta}\,^{\rho}\,_{\mu}\,^{\sigma}+C_{\alpha\rho\mu\sigma}C_{\beta}\,^{\rho}\,_{\lambda}\,^{\sigma}-\frac{1}{2}g_{\alpha\beta}C_{\rho\nu\lambda\sigma}C^{\rho\nu}\,_{\mu}\,^{\sigma}\nonumber \\
 &  & -\frac{1}{2}g_{\lambda\mu}C_{\alpha\rho\sigma\nu}C_{\beta}\,^{\rho\sigma\nu}+\frac{1}{8}g_{\alpha\beta}g_{\lambda\mu}C_{\rho\nu\sigma\eta}C^{\rho\nu\sigma\eta},\label{eq:Bel-Robinson}
\end{eqnarray}
and one has 
\begin{equation}
g^{\lambda\mu}\mathcal{B}_{\alpha\beta\lambda\mu}=\frac{D-4}{2}\mathcal{L}_{\alpha\beta}.
\end{equation}

To summarize, in this Section, we have shown that there is a part
of the Gauss-Bonnet tensor ${\cal H}_{\mu\nu}$ which is always higher
dimensional even though one part of the tensor can be made finite
with the procedure of dividing by $1/\left(D-4\right)$ and then formally
assuming that the remaining indices are four dimensional. The all
important point here is that if one bluntly drops the extra dimensional
part (which we called $\mathcal{L}_{\mu\nu}$), then the Bianchi Identity
is not satisfied for the remaining four-dimensional theory. Thus,
one either has all the dimensions, or one has 4 dimensions without
the benefit of the Bianchi identity. If one chooses the second option,
one cannot couple the four-dimensonal theory to conserved matter fields;
or one must impose the Bianchi identity on-shell for the solutions.

\section{The Field Equations In F\i rst-Order Formulation}

The authors of \cite{GlavanLin} argued that in the first-order formulation
of the Gauss-Bonnet theory a $\left(D-4\right)$ factor arises in
the field equations, and this factor can be canceled by introducing
the $\alpha/\left(D-4\right)$ factor in the action. This claim needs
to be scrutinized carefully as we do here. Let us just consider the
Gauss-Bonnet part of the action without any factors or coefficients.
Then, we have the $D$-dimensional action in terms of the vielbein
1-form $e^{a}$ and the curvature 2-form $R^{ab}:=d\omega^{ab}+\omega^{ac}\wedge\omega_{c}^{b}$;
\begin{align}
I_{GB} & =\int_{{\mathcal{M}}_{D}}\epsilon_{a_{1}a_{2}...a_{D}}R^{a_{1}a_{2}}\wedge R^{a_{3}a_{4}}\wedge e^{a_{5}}\wedge e^{a_{6}}...\wedge e^{a_{D}},
\end{align}
where the Latin indices refer to the tangent frame. Then, varying
the action with respect to the spin connection yields zero in the
zero torsion case; and the rest of the field equations are obtained
by varying with respect to the vielbein. At this stage the discussion
bifurcates\footnote{For $D\leq3$ the action vanishes identically and no further discussion
is needed.}: Assume that $D=4$, then the action reduces to $\int_{{\mathcal{M}}_{4}}\epsilon_{a_{1}a_{2}a_{3}a_{4}}R^{a_{1}a_{2}}\wedge R^{a_{3}a_{4}}$
where there is no vierbein left and one has 
\begin{equation}
\delta_{e^{a}}\int_{{\mathcal{M}}_{4}}\epsilon_{a_{1}a_{2}a_{3}a_{4}}R^{a_{1}a_{2}}\wedge R^{a_{3}a_{4}}=0,\qquad D=4.
\end{equation}
On the other hand, for generic $D>4$ dimensions, variation with respect
to the vielbein yields the field equation as a $\left(D-1\right)$-form
\begin{equation}
{\mathcal{E}}{a_{D}}=(D-4)\epsilon_{a_{1}a_{2}...a_{D}}R^{a_{1}a_{2}}\wedge R^{a_{3}a_{4}}\wedge e^{a_{5}}\wedge e^{a_{6}}...\wedge e^{a_{D-1}}\qquad D>4.
\end{equation}
Clearly the $\left(D-4\right)$ factor arises, but it does so \textit{only}
in $D$ dimensions: one cannot simply multiply with a $\alpha/\left(D-4\right)$
and take the $D\rightarrow4$ limit! In fact, starting from the last
equation, one can get the second order, metric form of the Gauss-Bonnet
tensor $\mathcal{H}_{\mu\nu}$, and in the process, one sees the role
played by this $\left(D-4\right)$ factor. To do so, instead of the
tangent frame indices we can recast it in terms of the spacetime indices
as which can be written as 
\begin{equation}
{\mathcal{E}}_{\nu}=\frac{(D-4)}{4}\epsilon_{\mu_{1}\mu_{2}...\mu_{D-1}\nu}R_{\phantom{\mu_{1}\mu_{2}}\sigma_{1}\sigma_{2}}^{\mu_{1}\mu_{2}}R_{\phantom{\mu_{3}\mu_{4}}\sigma_{3}\sigma_{4}}^{\mu_{3}\mu_{4}}dx^{\sigma_{1}}...\wedge dx^{\sigma_{4}}\wedge dx^{\mu_{5}}...\wedge dx^{\mu_{D-1}}.
\end{equation}
This is really a covariant vector-valued $\left(D-1\right)$-form,
and the Hodge dual of this $\left(D-1\right)$-form is a 1-form; and
since we have 
\begin{equation}
*\left(dx^{\sigma_{1}}...\wedge dx^{\sigma_{4}}\wedge dx^{\mu_{5}}...\wedge dx^{\mu_{D-1}}\right)=\epsilon^{\sigma_{1}...\sigma_{4}\mu_{5}...\mu_{D-1}}\,_{\mu_{D}}dx^{\mu_{D}},
\end{equation}
the 1-form field equations read 
\begin{equation}
*{\mathcal{E}}_{\nu}=\frac{\left(D-4\right)}{4}\epsilon_{\mu_{1}\mu_{2}...\mu_{D-1}\nu}\epsilon^{\sigma_{1}...\sigma_{4}\mu_{5}...\mu_{D-1}}\,_{\mu_{D}}R_{\phantom{\mu_{1}\mu_{2}}\sigma_{1}\sigma_{2}}^{\mu_{1}\mu_{2}}R_{\phantom{\mu_{3}\mu_{4}}\sigma_{3}\sigma_{4}}^{\mu_{3}\mu_{4}}dx^{\mu_{D}},
\end{equation}
from which we define the rank-2 tensor ${\mathcal{E}}_{\nu\alpha}$
as 
\begin{equation}
*{\mathcal{E}}_{\nu}=:{\mathcal{E}}_{\nu\alpha}dx^{\alpha}.
\end{equation}
Explicitly one has 
\begin{equation}
{\mathcal{E}}_{\nu\alpha}=\frac{\left(D-4\right)}{4}\epsilon_{\mu_{1}\mu_{2}...\mu_{D-1}\nu}\epsilon^{\sigma_{1}...\sigma_{4}\mu_{5}...\mu_{D-1}}\,_{\alpha}R_{\phantom{\mu_{1}\mu_{2}}\sigma_{1}\sigma_{2}}^{\mu_{1}\mu_{2}}R_{\phantom{\mu_{3}\mu_{4}}\sigma_{3}\sigma_{4}}^{\mu_{3}\mu_{4}},
\end{equation}
which can be further reduced with the help of the identity 
\begin{equation}
\epsilon_{\mu_{1}\mu_{2}...\mu_{D-1}\nu}\epsilon^{\sigma_{1}...\sigma_{4}\mu_{5}...\mu_{D-1}}\,_{\alpha}=-\left(D-5\right)!g_{\beta\alpha}\delta_{\mu_{1}...\mu_{4}\nu}^{\sigma_{1}..\sigma_{4}\beta},
\end{equation}
where we used the generalized Kronecker delta. So, we have 
\begin{eqnarray}
{\mathcal{E}}_{\nu\alpha} & = & -\frac{\left(D-4\right)}{4}\left(D-5\right)!g_{\beta\alpha}\delta_{\mu_{1}...\mu_{4}\nu}^{\sigma_{1}..\sigma_{4}\beta}R_{\phantom{\mu_{1}\mu_{2}}\sigma_{1}\sigma_{2}}^{\mu_{1}\mu_{2}}R_{\phantom{\mu_{3}\mu_{4}}\sigma_{3}\sigma_{4}}^{\mu_{3}\mu_{4}}\nonumber \\
 &  & =-\frac{\left(D-4\right)!}{4}g_{\beta\alpha}\delta_{\mu_{1}...\mu_{4}\nu}^{\sigma_{1}..\sigma_{4}\beta}R_{\phantom{\mu_{1}\mu_{2}}\sigma_{1}\sigma_{2}}^{\mu_{1}\mu_{2}}R_{\phantom{\mu_{3}\mu_{4}}\sigma_{3}\sigma_{4}}^{\mu_{3}\mu_{4}}.
\end{eqnarray}
Observe that the $\left(D-4\right)$ factor turned into $\left(D-4\right)!$
which does not vanish for $D=4$. Since one also has 
\begin{equation}
g_{\beta\alpha}\delta_{\mu_{1}...\mu_{4}\nu}^{\sigma_{1}..\sigma_{4}\beta}R_{\phantom{\mu_{1}\mu_{2}}\sigma_{1}\sigma_{2}}^{\mu_{1}\mu_{2}}R_{\phantom{\mu_{3}\mu_{4}}\sigma_{3}\sigma_{4}}^{\mu_{3}\mu_{4}}=-8{\mathcal{H}}_{\nu\alpha},
\end{equation}
where ${\mathcal{H}}_{\nu\alpha}$ is the Gauss-Bonnet tensor we defined
above, we get 
\begin{equation}
{\mathcal{E}}_{\nu\mu}=2\left(D-4\right)!{\mathcal{H}}_{\nu\alpha}
\end{equation}
Thus, the moral of the story is that one either has an explicit $\left(D-4\right)$
factor in front of the field equations when they are written in terms
of the vielbeins and the spin connection where the dimensionality
of the spacetime is explicitly $D>4$ as counted by the number of
vielbeins; or, one does not have an explicit $\left(D-4\right)$ factor
in the field equations in the metric formulation. There is no other
option. In the metric formulation, we have shown in the previous section
that a $\left(D-4\right)$ does not arise for generic metrics in all
parts of the field equations.

\section{Direct-Product Spacetimes}

To see the alluded problems explicitly in an example, let us consider
the direct-product spacetimes for which the $D$-dimensional metric
has the form 
\begin{equation}
ds^{2}=g_{AB}dx^{A}dx^{B}=g_{\alpha\beta}\left(x^{\mu}\right)dx^{\alpha}dx^{\beta}+g_{ab}\left(x^{c}\right)dx^{a}dx^{b},
\end{equation}
where $A,B=1,2\cdots,D$; $\alpha,\beta=1,2,3,4$; and $a,b=5,6,\cdots,D$.
Here, $g_{\alpha\beta}$ depends only on the four-dimensional coordinates
$x^{\mu}$, and $g_{ab}$ depends only on the extra dimensional coordinates
$x^{c}$. Then, for the Christoffel connection, 
\begin{equation}
\Gamma_{BC}^{A}=\frac{1}{2}g^{AE}\left(\partial_{B}g_{EC}+\partial_{C}g_{EB}-\partial_{E}g_{BC}\right),
\end{equation}
it is easy to show that the only nonzero parts are 
\begin{align}
_{D}\Gamma_{\beta\mu}^{\alpha}= & _{4}\Gamma_{\beta\mu}^{\alpha}=\frac{1}{2}g^{\alpha\epsilon}\left(\partial_{\beta}g_{\mu\epsilon}+\partial_{\mu}g_{\beta\epsilon}-\partial_{\epsilon}g_{\beta\mu}\right),\\
_{D}\Gamma_{bc}^{a}= & _{d}\Gamma_{bc}^{a}=\frac{1}{2}g^{ae}\left(\partial_{b}g_{ce}+\partial_{c}g_{be}-\partial_{e}g_{bc}\right),
\end{align}
where the subindex $d$ denotes the $\left(D-4\right)$-dimensional.
Due to this property, we have the following nonzero components of
the Riemann tensor, $R_{\phantom{A}BCE}^{A}$, and Ricci tensor, $R_{AB}$;
\begin{align}
_{D}R_{\phantom{\alpha}\beta\mu\epsilon}^{\alpha} & =_{4}R_{\phantom{\alpha}\beta\mu\epsilon}^{\alpha},\qquad{}_{D}R_{\phantom{a}bce}^{a}=_{d}R_{\phantom{a}bce}^{a},\nonumber \\
_{D}R_{\alpha\beta} & =_{4}R_{\alpha\beta},\qquad{}_{D}R_{ab}=_{d}R_{ab}.
\end{align}
In addition, the scalar curvature splits as 
\begin{align}
_{D}R= & _{4}R+\,_{d}R.
\end{align}
The nonzero components of the Weyl tensor $C_{ABEF}$ are

\noindent 
\begin{align}
_{D}C_{\alpha\beta\epsilon\nu}= & _{4}C_{\alpha\beta\epsilon\nu}+\frac{\left(D-4\right)}{\left(D-2\right)}\left(g_{\alpha[\epsilon}\,_{4}R_{\nu]\beta}-g_{\beta[\epsilon}\,_{4}R_{\nu]\alpha}\right)\nonumber \\
 & -\frac{\left(D-4\right)\left(D+1\right)}{3\left(D-1\right)\left(D-2\right)}{}_{4}Rg_{\alpha[\epsilon}g_{\nu]\beta}+\frac{2}{\left(D-1\right)\left(D-2\right)}\,_{d}Rg_{\alpha[\epsilon}g_{\nu]\beta},\label{eq:Weyl_4}\\
_{D}C_{abef}= & _{d}C_{abef}+\frac{8}{\left(D-2\right)\left(D-6\right)}\left(g_{a[e}\,_{d}R_{f]b}-g_{b[e}\,_{d}R_{f]a}\right)\nonumber \\
 & -\frac{8\left(2D-7\right)}{\left(D-1\right)\left(D-2\right)\left(D-5\right)\left(D-6\right)}\,_{d}Rg_{a[e}g_{f]b}+\frac{2}{\left(D-1\right)\left(D-2\right)}{}_{4}Rg_{a[e}g_{f]b},\label{eq:Weyl_d}\\
_{D}C_{\alpha ba\beta}= & \frac{1}{\left(D-2\right)}\left(g_{\alpha\beta}\,_{d}R_{ab}+g_{ab}\,_{4}R_{\alpha\beta}\right)-\frac{1}{\left(D-1\right)\left(D-2\right)}\left(_{4}R+\,_{d}R\right)g_{\alpha\beta}g_{ab},\label{eq:Weyl_mixed}
\end{align}
in addition to $_{D}C_{\alpha b\beta a}=-{}_{D}C_{b\alpha\beta a}={}_{D}C_{b\alpha a\beta}=-{}_{D}C_{\alpha ba\beta}$.

If the $d$-dimensional internal space is flat as 
\begin{equation}
ds^{2}=g_{AB}dx^{A}dx^{B}=g_{\alpha\beta}\left(x^{\mu}\right)dx^{\alpha}dx^{\beta}+\eta_{ab}dx^{a}dx^{b},
\end{equation}
then one has 
\begin{equation}
_{D}R_{\phantom{a}bce}^{a}=0,\qquad{}_{D}R_{ab}=0,\qquad{}_{D}R={}_{4}R,
\end{equation}
and nonzero components of the Weyl tensor given in (\ref{eq:Weyl_4}-\ref{eq:Weyl_mixed})
become 
\begin{align}
_{D}C_{\alpha\beta\epsilon\nu}= & _{4}C_{\alpha\beta\epsilon\nu}+\frac{\left(D-4\right)}{\left(D-2\right)}\left(g_{\alpha[\epsilon}\,_{4}R_{\nu]\beta}-g_{\beta[\epsilon}\,_{4}R_{\nu]\alpha}\right)-\frac{\left(D-4\right)\left(D+1\right)}{3\left(D-1\right)\left(D-2\right)}{}_{4}Rg_{\alpha[\epsilon}g_{\nu]\beta},\label{eq:Weyl_4-flat}\\
_{D}C_{abef}= & \frac{2}{\left(D-1\right)\left(D-2\right)}{}_{4}R\eta_{a[e}\eta_{f]b},\label{eq:Weyl_d-flat}\\
_{D}C_{\alpha ba\beta}= & \frac{1}{\left(D-2\right)}\eta_{ab}\,_{4}R_{\alpha\beta}-\frac{1}{\left(D-1\right)\left(D-2\right)}{}_{4}Rg_{\alpha\beta}\eta_{ab},\label{eq:Weyl_mixed-flat}
\end{align}
in addition to $_{D}C_{\alpha b\beta a}=-{}_{D}C_{b\alpha\beta a}={}_{D}C_{b\alpha a\beta}=-{}_{D}C_{\alpha ba\beta}$.

With the above results, let us provide a clear example of where the
limit 
\begin{equation}
\lim_{D\rightarrow4}\mathcal{L}_{AB}=\lim_{D\rightarrow4}\left[\frac{1}{D-4}\left(C_{AEFG}C_{B}^{\phantom{B}EFG}-\frac{1}{4}g_{AB}C_{EFGH}C^{EFGH}\right)\right],
\end{equation}
fails. Consider the $\mathcal{L}_{ab}$ components of the Lanczos-Bach
tensor, 
\begin{equation}
\mathcal{L}_{ab}=C_{aEFG}C_{b}^{\phantom{b}EFG}-\frac{1}{4}\eta_{ab}C_{EFGH}C^{EFGH}.
\end{equation}
These components can be written as 
\begin{align}
\mathcal{L}_{ab}= & \left(\,_{D}C_{aefg}\,_{D}C_{b}^{\phantom{}efg}+2\,_{D}C_{a\epsilon f\gamma}\,_{D}C_{b}^{\phantom{b}\epsilon f\gamma}\right)\nonumber \\
 & -\frac{1}{4}\eta_{ab}\left(\,_{D}C_{\alpha\epsilon\nu\gamma}\,_{D}C^{\alpha\epsilon\nu\gamma}+\,_{D}C_{aefg}\,_{D}C^{aefg}+4\,_{D}C_{\alpha e\gamma f}\,_{D}C^{\alpha e\gamma f}\right).
\end{align}
By using (\ref{eq:Weyl_4-flat}-\ref{eq:Weyl_mixed-flat}), the $\mathcal{L}_{ab}$
components of the Lanczos-Bach tensor can be calculated in terms of
the four-dimensional and $d$-dimensional quantities as 
\begin{equation}
\mathcal{L}_{ab}=-\frac{1}{4}\eta_{ab}\left(\,_{4}C_{\epsilon\nu\gamma\eta}\,_{4}C^{\epsilon\nu\gamma\eta}+\frac{2\left(D^{2}-6D+4\right)}{\left(D-2\right)^{2}}\,_{4}R_{\epsilon\nu}\,_{4}R^{\epsilon\nu}-\frac{\left(D^{3}-5D^{2}+2D-16\right)}{3\left(D-1\right)\left(D-2\right)^{2}}\,_{4}R^{2}\right).
\end{equation}
Then, the $D\rightarrow4$ limit for this term in the form, 
\begin{equation}
\lim_{D\rightarrow4}\left[-\frac{\eta_{ab}}{4\left(D-4\right)}\left(\,_{4}C_{\epsilon\nu\gamma\eta}\,_{4}C^{\epsilon\nu\gamma\eta}+\frac{2\left(D^{2}-6D+4\right)}{\left(D-2\right)^{2}}\,_{4}R_{\epsilon\nu}\,_{4}R^{\epsilon\nu}-\frac{\left(D^{3}-5D^{2}+2D-16\right)}{3\left(D-1\right)\left(D-2\right)^{2}}\,_{4}R^{2}\right)\right],
\end{equation}
is undefined, and this fact indicates that in general, there is no
proper $D\rightarrow4$ limit for the field equations for the direct-product
spacetimes.

\section{Conclusions}

Recently \cite{GlavanLin}, contrary to the common knowledge and to
the Lovelock's theorem \cite{Lovelock1,Lovelock2,Lanczos}, a novel
\textit{four-dimensional } Einstein-Gauss-Bonnet theory with only
a massless spin-2 graviton degree of freedom was suggested to exist.
A four-dimensional gravity theory should have four-dimensional equations:
here, we have shown that this is not the case. Namely, we have shown
that the novel Einstein-Gauss-Bonnet theory in four dimensions does
not have an intrinsically four-dimensional description in terms of
a covariantly-conserved rank-2 tensor in four dimensions. We have
done this by splitting the Gauss-Bonnet tensor (\ref{eq:EGB}) into
two parts as (\ref{eq:H_split}): one is what we called the Lanczos-Bach
tensor (\ref{eq:L}) which is related to the trace of the $D$-dimensional
Bel-Robinson tensor which does not have an explicit $\left(D-4\right)$
factor, and the other part (\ref{eq:Z}) is a part that has an explicit
$\left(D-4\right)$ factor in front. The Lanczos-Bach tensor vanishes
identically in four dimensions; however, it cannot be set to identically
zero in that dimensions since in the absence of it, the Gauss-Bonnet
tensor does not satisfy the Bianchi Identity. Thus, the theory must
be defined in $D>4$ dimensions to be nontrivial which is in complete
agreement with the Lovelock's theorem. But, once the theory is defined
in $D$ dimensions, it will have all sorts of $D$ dimensional solutions
and in none of these solutions one can simply dispose of $\left(D-4\right)$
dimensions or coordinates as such a discrimination among spacetime
dimensions simply does not make sense. We gave an explicit example
in the form of a direct product. In the first-order formulation with
the vielbein and the spin connection, there is an explicit $\left(D-4\right)$
factor in front of the field equations, but this factor only arises
in $D>4$ dimensions not in four dimensions. What we have shown here
for the Gauss-Bonnet tensor in its critical $D=4$ dimensions is also
valid for the other Lovelock tensors in their critical dimensions.

\section{Acknowledgment}

We would like to thank S.~Deser and Y.~Pang for useful discussions.

\end{document}